\documentclass[12pt,preprint]{aastex}
\usepackage{amsmath,amsfonts}
\begin{document}
\title{Masses and orbital inclinations of planets in the PSR~B1257+12 system}
\author{Maciej Konacki \altaffilmark{1,2}} \and \author{Alex Wolszczan \altaffilmark{3,4}}
\altaffiltext{1}{Department of Geological and Planetary Sciences, California
Institute of Technology, MS 150-21, Pasadena, CA 91125, USA; 
e-mail: maciej@gps.caltech.edu}
\altaffiltext{2}{Nicolaus Copernicus Astronomical Center, Polish Academy of Sciences,
Rabia\'nska 8, 87-100 Toru\'n, Poland}
\altaffiltext{3}{Department of Astronomy and Astrophysics, Penn State University
University Park, PA 16802, USA; e-mail: alex@astro.psu.edu}
\altaffiltext{4}{Toru\'n Centre for Astronomy, Nicolaus Copernicus University
ul. Gagarina 11, 87-100 Toru\' n, Poland}
\begin{abstract}
We present measurements of the true masses and orbital inclinations of the two
Earth-mass planets in the PSR B1257+12 system, based on the analysis of
their mutual gravitational perturbations detectable as microsecond variations
of the arrival times of radio pulses from the pulsar. The 6.2-millisecond 
pulsar, PSR B1257+12, has been regularly timed with the Arecibo telescope 
since late 1990. Assuming the standard pulsar mass of 1.4 $M_{\odot}$, 
the derived masses of planets B and C are 4.3$\pm$0.2$M_{\oplus}$ 
and 3.9$\pm$0.2$M_{\oplus}$, respectively. The corresponding orbital inclinations 
of 53$^{\circ}\pm$4$^{\circ}$ and 47$^{\circ}\pm$3$^{\circ}$ (or 127$^{\circ}$ and 
133$^{\circ}$) imply that the two orbits are almost coplanar. This result, together 
with the known near 3:2 resonance between the orbits of the two planets, strongly
supports the hypothesis of a disk origin of the PSR B1257+12 planetary system.
The system's long-term stability is guaranteed by the low, Earth-like masses of 
planets B and C.

\end{abstract}
\keywords{planetary systems --- pulsars: individual (PSR~B1257+12)}

\section{Introduction}
The first extrasolar planetary system consisting of three planets orbiting
a neutron star, the 6.2-ms radio pulsar PSR B1257+12, has been systematically
observed since the time of its discovery 
\citep{Wol:92::,Wol:94::,Wol:00::}. 
Pulsar planets comprise the only known system of terrestrial-mass planets beyond
the Sun. They exhibit a striking dynamical similarity to the inner solar
system \citep{Maz:95::}. The pulsar planets have also provided the first
demonstration of orbital near resonance in an extrasolar planetary system 
with the resulting, measurable gravitational
perturbations \citep{Wol:94::,Kon:99::}. In the recent years, a variety
of resonances and commensurabilities have been detected in
planetary systems around normal stars \citep{Nel:02::,Fis:03::}.
A microsecond precision of the pulsar timing offers a possibility
to detect additional, even lower mass planets around PSR B1257+12
and other neutron stars observable as radio pulsars \citep{Wol:97::} 

An accurate knowledge of masses and orbital inclinations of
the PSR B1257+12 planets is critically important for our understanding of
the origin and evolution of this unique planetary system. In this case,
the near 3:2 mean motion resonance (MMR) between planets B and C in the 
pulsar system and the existence of detectable gravitational perturbations 
between the two planets \citep{Ras:92::,Mal:92::,Pea:93::,Wol:94::,Kon:99::}
provide the mechanism to derive their masses without an a priori knowledge 
of orbital inclinations. An approximate analytical model which 
includes the effect of gravitational interactions between planets B and C
has been published by \cite{Mal:93::}.
\cite{Kon:00::} have developed a new semi-analytical model in which 
perturbations between the two planets are parametrized in terms of 
the two planetary masses and the mutual orientation of the orbits with
a sufficient precision to make a practical application of this approach feasible. 
Using the simulated data, they have demonstrated that the planet
masses and hence their orbital inclinations can be derived from a least-squares
fit of this model to the pulse times-of-arrival (TOA) measurements spanning
a sufficiently long period of time.

In this paper, we describe the results of putting this model to a practical test
by applying it to the TOA measurements of PSR B1257+12 made with the 305-m 
Arecibo
telescope over a 12-year period between 1990 and 2003. In Section 2, we present
a brief summary of the timing model published by \cite{Kon:00::}.
Section 3 contains the description of Arecibo timing measurements of PSR 
B1257+12
and the details and results of data analysis. Consequences of our successful
determination of masses and orbital inclinations of planets B and C in the
PSR B1257+12 system and future prospects for new pulsar planet detections 
are discussed in Section 4.

\section{The timing model}

As the first approximation, the orbits of the PSR B1257+12 system
can be described in terms of the sum of the Keplerian motions of its planets,
in which case the direct and indirect gravitational interactions between
planets are negligible. However, as predicted by
\cite{Ras:92::}, \cite{Mal:92::}, and \cite{Pea:93::}, the 3:2 commensurability 
of the orbital periods of planets B and C does lead to observable deviations 
from the simple Keplerian model \citep{Wol:94::,Kon:99::}. 

In the classical approach, such departures from the Keplerian dynamics 
can be described by means of the so-called osculating 
orbital elements by invoking the Keplerian orbital elements that are no longer 
constant but they change in time due to mutual interactions between planets. 
For PSR B1257+12 timing,
the difficulty with this approach is that only a radial component of the
spatial motion of the pulsar can be measured
and a complete information on the orientation of orbits
of the planets is not available.
Below, we summarize an approach developed by \cite{Kon:00::} that addresses
this problem and allows to accurately calculate the osculating elements
of planetary orbits in the presence of gravitational perturbations.

In the new timing model, we define the TOA variations $\Delta\tau(t)$ as a sum
of the two components
\begin{equation}
\Delta\tau(t) = \Delta\tau_{kep}(x_j^0,e_j^0,\omega_j^0,P_j^0,T_{pj}^0,t) 
+ \delta\tau_{int}(\Delta x_j,\Delta e_j,\Delta \omega_j,\Delta P_j,
\Delta T_{pj},t)
\end{equation}
Here, $\Delta\tau_{kep}$ describes the TOA variations due to the 
Keplerian part of the motion and is a function of the instantaneous values of 
the Keplerian elements of planets at the moment $t_0$. For a $j$-th planet, these are
the projected semi-major axis of the pulsar orbit $x_j^0$, eccentricity
$e_j^0$, longitude of the periastron $\omega_j^0$, orbital period $P_j^0$, and time of 
the periastron passage $T_{pj}^0$, respectively. The second term in Eqn. (1),
$\delta\tau_{int}$, describes the TOA variations caused by
changes in the osculating orbital elements, $\Delta x_j,\Delta e_j,
\Delta \omega_j,\Delta P_j, \Delta T_{pj}$. These variables are functions of 
masses of the pulsar and the planets and of the relative geometry of the orbits.

In the model, the masses are expressed in terms of two parameters, 
$\gamma_B = m_B/M_{psr}$, and $\gamma_C =  m_C/M_{psr}$, 
where $m_B,m_C,M_{psr}$ are the masses of the two planets and the pulsar, respectively.
As shown in \cite{Kon:00::}, geometry of the orbits can be
described in terms of a relative position of the node of the orbits, $\tau$,
if their relative inclination is small ($I \le 10^{\circ}$).
The parameter $\tau$ is related to 
the longitudes of the ascending nodes,  $\Omega_B,\Omega_C$, and the orbital 
inclinations $i_B,i_C$ through the following set of equations 
\citep[see also Fig.~1 in][]{Kon:00::}
\begin{equation}
\label{tau}
\begin{array}{l}
\cos (I/2) \sin (\tau/2) = \sin ((\Omega_C - \Omega_B)/2) \cos ((i_C + i_B)/2),\\
\cos (I/2) \cos (\tau/2) = \cos ((\Omega_C - \Omega_B)/2) \cos ((i_C - i_B)/2)
\end{array}
\end{equation}
As long as the above small-angle approvimation is valid,
the parameters $\gamma_B,\gamma_C$, and $\tau$ represent an accurate description
of the time evolution 
of the osculating orbital elements. Consequently, Eqn. (1) can be rewritten as 
\begin{equation}
\Delta\tau(t) = \Delta\tau_{kep}(x_j^0,e_j^0,\omega_j^0,P_j^0,T_{pj}^0,t) 
+ \delta\tau_{int}(\gamma_B,\gamma_C,\tau,t)
\end{equation}
to define, in general terms, a modified timing model which includes familiar quantities,
${x_j^0,e_j^0,\omega_j^0,P_j^0,T_{pj}^0}$, to parametrize Keplerian orbits 
and introduces three additional parameters,
$\gamma_B$, $\gamma_C$, and $\tau$, to account for
the perturbations between planets B and C.
The correctness of this approach has been verified by extensive simulations
described by \cite{Kon:00::}.

\section{Observations and data analysis}

PSR B1257+12 has been observed with the Arecibo telescope since its
discovery in 1990 \citep{Wol:90::}. In the years 1990-1994, before
the Arecibo upgrade, the pulsar had been timed at 430 MHz and 1400 MHz 
with the Princeton Mark-III backend that utilizes two 32-channel 
filterbanks \citep[for a description see][]{Stin:92::}. Starting in 1994, 
just before the beginning of the Arecibo upgrade, the pulsar had also been timed 
with a 128-channel filterbank-based Penn State Pulsar 
Machine \cite[PSPM;][]{Cad:97::} at 430 MHz. These observations were very useful
in determining the timing offsets between the Mark-III and the PSPM data sets 
and
in eliminating another offset that was found between the data acquired with the PSPM 
before and
after the Arecibo upgrade.
Observations with the PSPM were resumed after the upgrade in November 1997.
Since then, the pulsar has been systematically timed 
at 430 MHZ and 1400 MHz at 3-4 week intervals. 
A more detailed description of the data acquisition and the TOA measurement process
can be found in \cite{Wol:00::}. 

The timing model 
included the pulsar spin and astrometric parameters, Keplerian elements of the 
orbits
of planets A, B, and C, and the three variables $\gamma_B$, $\gamma_C$, and 
$\tau$
introduced to parametrize perturbations between planets B and C, as described 
above. 
A propagation delay and its long-term decline
due to the varying line-of-sight electron density were parametrized
in terms of the dispersion measure (DM) and its first three time derivatives. 
Low-amplitude DM variations on the timescales of hundreds of days have been 
removed
by means of direct measurements of local DM values averaged over consecutive 3 month 
intervals.
The most recent version of the timing analysis package TEMPO
(see {\tt http://pulsar.princeton.edu/tempo}), was modified to incorporate
this model and to least-squares fit it
to the observed topocentric TOAs.
The final best-fit residuals for 
daily-averaged
TOAs are characterized by a $\sim$3.0 $\mu$s rms noise which is consistent with
a predicted value of $\sim$2.0 $\mu$s based on the observing parameters and the system performance
\citep[e.g.][]{Fos:90::}.
The residuals for three fits to data involving 
different sets of parameters are shown in Fig. 1 and the model parameters 
for the final fit of the full timing model are listed in Tables 1 and 2.

The new timing model for PSR B1257+12 offers further improvement of the accuracy
of the determination of the standard pulsar and planetary parameters and, most 
importantly, it includes highly significant values for the three perturbation-related 
parameters, $\gamma_B$, $\gamma_C$, and $\tau$ (Fig. 2).
From $m_B=\gamma_B M_{psr},m_C=\gamma_C M_{psr}$, one obtains the masses of planets B and C to be
4.3$\pm$0.2$M_{\oplus}$ and 3.9$\pm$0.2$M_{\oplus}$, respectively, using the canonical
pulsar mass, $M_{psr}$ = 1.4 $M_{\odot}$. Since the scatter in the measured neutron star
masses is small \citep{Tho:99::}, it is unlikely that a possible error
in the assumed pulsar mass would significantly affect these results.
Because of the $\sin(i)$ ambiguity, there are four possible
sets of the orbital inclinations for the planets B and C: $(53^{\circ},47^{\circ})$,
$(127^{\circ},133^{\circ})$ corresponding to the difference in the ascending
nodes $\Omega_C-\Omega_B\approx0^{\circ}$ (relative inclination $I\approx6^{\circ}$),
and $(53^{\circ},133^{\circ})$, $(127^{\circ},47^{\circ})$, corresponding to the 
difference in the ascending nodes $\Omega_C-\Omega_B\approx180^{\circ}$ 
(relative inclination $I\approx174^{\circ}$). 
Obviously, in both cases the planets have nearly coplanar orbits, but in the
latter one, their orbital motions have opposite senses.
Because our numerical simulations of the system's dynamics
show that this situation leads to distinctly different perturbative TOA 
variations that are not observed, 
only the first two sets of the orbital inclinations, 
53$^{\circ}\pm$4$^{\circ}$ and 47$^{\circ}\pm$3$^{\circ}$ or 127$^{\circ}$ 
and 133$^{\circ}$ are plausible. This implies that the two planets move in
nearly coplanar orbits in the same sense. In addition, with the known value of 
$\tau$ (Table~2), 
one obtains $\Omega_C-\Omega_B\approx3^{\circ}$ or 
$\Omega_C-\Omega_B\approx-3^{\circ}$ from equation \eqref{tau}. 
Since it is reasonable to assume that the 
inner planet A is in the same plane, its mass given in Table 2 has been calculated 
for orbital inclination of 50$^{\circ}$. Although the formal errors of the orbital
inclinations allow their relative inclination, $I$, to
be as high as $\sim\!13^{\circ}$, such a departure from the
model assumption of $I\leq 10^{\circ}$ would have little effect on the best-fit
masses of the planets \citep{Kon:00::}.

\section{Discussion}

The results described in this paper demonstrate that, under special circumstances 
created by the existence of measurable gravitational perturbations between planets 
B and C in the PSR~B1257+12 system, it is possible to determine their true masses and 
orbital inclinations. A near 3:2 MMR between the orbits of the 
two planets and the fact that they are nearly coplanar imply that the pulsar system 
has been created as the result of a disk evolution similar to that invoked to describe 
planet formation around normal stars \citep{Bos:03::}.
This represents a firm observational constraint which requires that any viable theory 
of the origin of the pulsar planets provides means to create a circumpulsar disk of 
matter that survives long enough and has a sufficient angular momentum to enable 
planet formation. Continuing timing observations of PSR B1257+12 will eventually 
settle the problem of a fourth, more distant planet (or planets) around it 
\citep{Wol:00::} and provide further constraints on the origin and evolution 
of this planetary system. 

Another important consequence of the determination of true masses of planets B 
and C is the implied long-term stability of the pulsar 
system. This problem has been investigated by \cite{Ras:92::} and \cite{Mal:92::}, 
who have established that the two planets would have to be as massive as 2-3 Jupiter 
masses to render the system dynamically unstable on a $10^4-10^5$ yr timescale. 
Obviously, the measured, terrestrial masses of the planets (Table 2) are much 
too low to create such a condition. In fact, this conclusion is not surprising, given
another result of
\cite{Mal:92::}, who have calculated that, if the
masses of the two planets were about 20-40$M_{\oplus}$, the
system would be locked in the exact 3:2 MMR and the character of 
perturbations would be very different from the observed near-resonance configuration.

The early theories of pulsar planet formation have been summarized by \cite{Pod:93::}
and further discussed by \cite{Phi:93::}. More recently, \cite{Mil:01::} 
and \cite{Han:02::} have examined the conditions of survival 
and evolution of pulsar protoplanetary disks. They have concluded that an initially 
sufficiently massive ($>10^{28}$g) disk would be able to resist evaporation by the 
pulsar accretion flux and create planets on a typical, $\sim 10^7$-year timescale. 
A quick formation of a massive disk around the pulsar could, for instance,
be accomplished by tidal disruption of a stellar companion \citep{Ste:92::,Phi:93::} 
or, possibly,
in the process of a white dwarf merger \citep{Pod:91::,Liv:92::}. 
Both these processes, although 
entirely feasible, cannot be very common. In fact, with the exception of PSR B1257+12, 
no planetary companions have emerged
from the precision timing of 48 galactic millisecond pulsars
\citep{Lor:01::},
implying their rarity, independently of the specific 
formation mechanism.

Since the current evidence points to isolated millisecond pulsars as best candidates 
for a presence of planetary companions around them \citep{Wol:97::,Mil:01::}, 
new detections of such objects 
by the ongoing and future pulsar surveys  will be very important. So far, only 10 
solitary millisecond pulsars, including PSR B1257+12, have been discovered.
This remains in a stark contrast with the sample of about 2000 solar-type 
stars that are included in the Doppler surveys for extrasolar planets \citep{Mar:03::}. 
Factors to be taken into account while designing pulsar planet searches 
must include the fact that such pulsars are less common and appear to be intrinsically
even fainter than more typical, binary milisecond pulsars \citep{Bai:97::}. 
In addition, if the high space velocity of PSR B1257+12 ($\sim$300 km s$^{-1}$, Table 1) 
and the fact that it has planets were causally connected, such objects would spend most of 
the time near turnover points of their galactic orbits, which would make them 
difficult to detect. Altogether, it appears that further improvement of the statistics
of neutron star planetary systems may be a lengthy process, even if they are similar to
those established for the occurence of giant planets around normal 
stars \citep{Mar:03::}. 

\acknowledgements

M.~K. is a Michelson Postdoctoral Fellow and is partially supported by
the Polish Committee for Scientific Research, Grant No.~2P03D~001~22.
A.~W. is supported by the NASA grant NAG5-4301 and by the NSF under grant AST-9988217.  
The authors thank the referee, Renu Malhotra, for insightful comments on the
manuscript.

\clearpage

%
% FIGURE CAPTIONS
%

\figcaption[f1.eps]{The best-fit, daily-averaged TOA residuals for three timing
models of PSR B1257+12 observed at 430 MHz. The filled circles and triangles
represent TOAs measured with the Mark-III and the PSPM backends, respectively. 
The solid line marks the predicted TOA variations for each timing model.
{\it (a)} TOA residuals after the fit of the standard timing model without planets.
TOA variations are dominated by the Keplerian orbital effects from planets B and C.
{\it (b)} TOA residuals for the model including the Keplerian orbits of planets A, B and C.
Residual variations are determined by perturbations between planets B and C.
{\it (c)} Residuals for the model including all the standard pulsar parameters and
the Keplerian and non-Keplerian orbital effects.}

\figcaption[f2.eps]{Determination of the non-Keplerian parameters 
describing perturbations between planets B and C
in the timing
model for PSR B1257+12. 
{\it (a)} $\Delta\chi^2 = \chi^2 - 
\chi^2_{min}$
as a function of $\tau$. 
{\it (b)} $\Delta\chi^2$ as a function of planet masses, $m_B$ and $m_C$, 
obtained from $m_B=\gamma_B M_{psr},m_C=\gamma_C M_{psr}$ for the pulsar mass 
$M_{psr} = 1.4 M_{\odot}$. Both functions exhibit well-defined minima.} 

\clearpage

%
% FIGURES   
%

%
% FIGURE 1
%

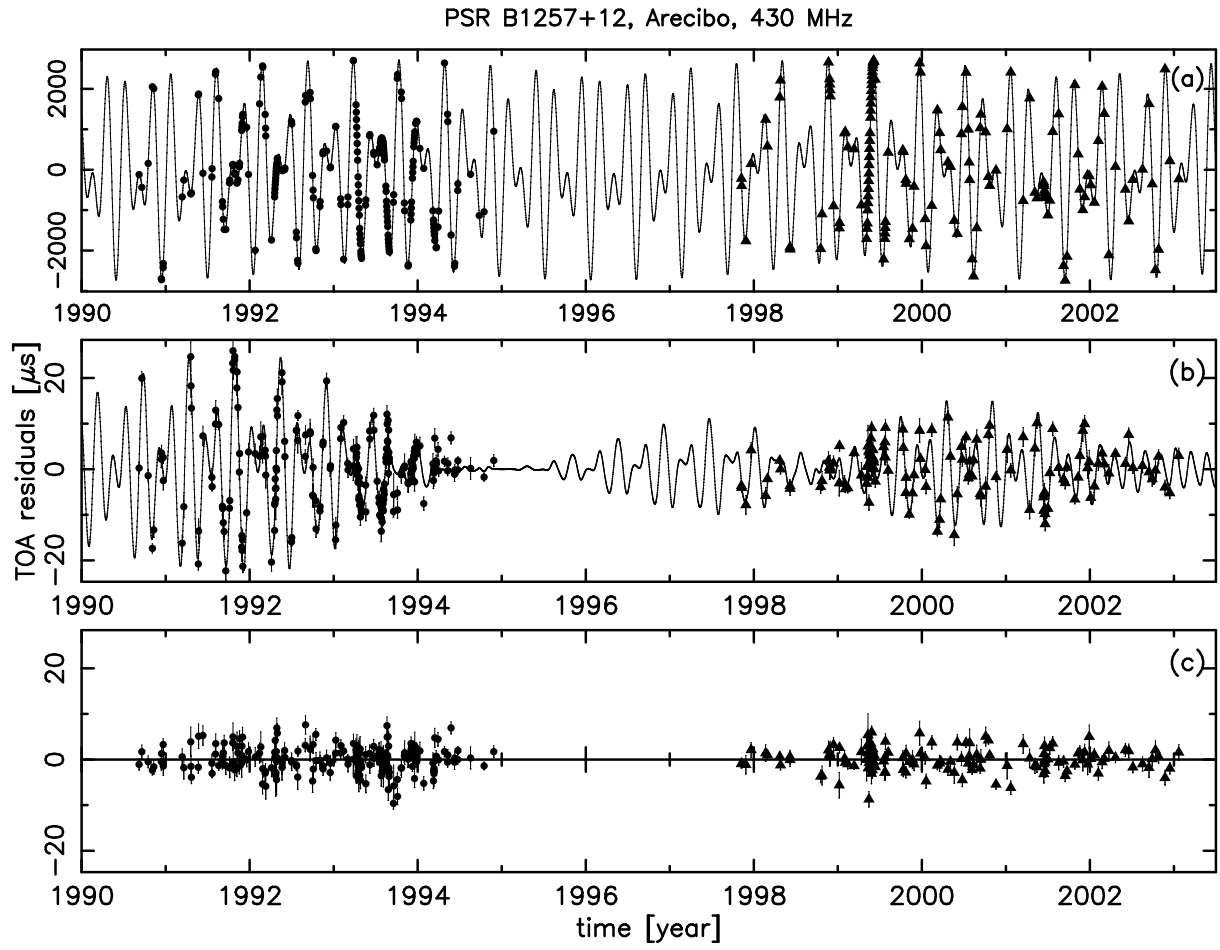
\begin{figure}
\figurenum{1}
\includegraphics[angle=-90,scale=0.65]{f1.eps}
\caption{}
\end{figure}

\clearpage

%
% FIGURE 2
%

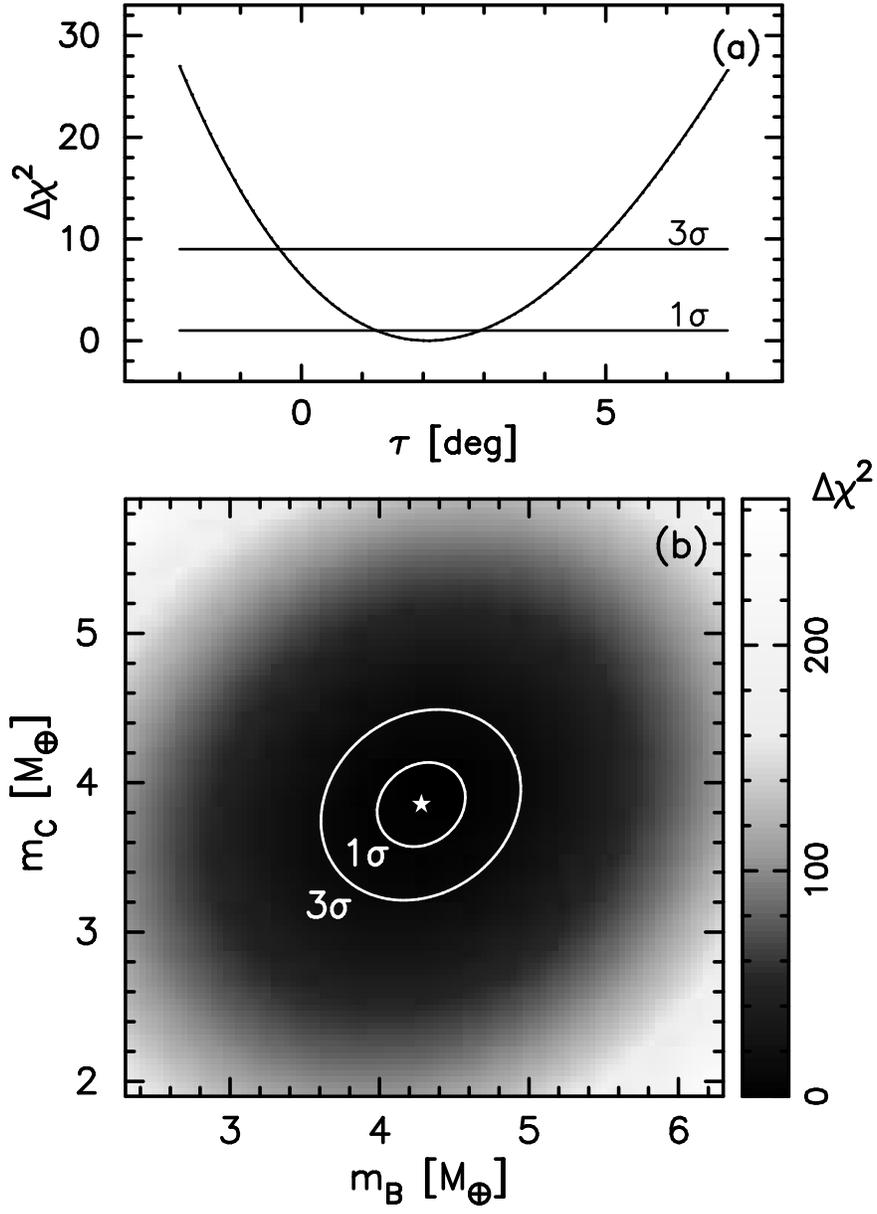
\begin{figure}
\figurenum{2}
\epsscale{0.7}
\plotone{f2.eps}
\caption{}
\end{figure}

\clearpage

%
% TABLES
%

%
% TABLE 1
%

\begin{deluxetable}{lc}
%\footnotesize
%\scriptsize
\tablewidth{315pt}
\tablecaption{Timing Parameters and Derived Quantities for PSR~B1257+12
\tablenotemark{a}}
\tablehead{\colhead{Parameter} & \colhead{PSR B1257+12} }
\startdata
Right ascension, $\alpha$ (J2000)\dotfill&  13$^h$00$^m$03.$\!\!^s$5767(1)\\     
Declination, $\delta$ (J2000)\dotfill& 
12$^\circ$40$^\prime$56.$\!\!^{\prime\prime}$4721(3)\\
Proper motion in $\alpha$, $\mu_{\alpha}$ (mas/yr)\dotfill & 45.50(4)\\
Proper motion in $\delta$, $\mu_{\delta}$ (mas/yr)\dotfill&   -84.70(7)\\
Period, $P$ (ms)\dotfill   &  6.21853194840048(3)\\              
Period derivative, $\dot P$ (10$^{-20}$)\dotfill &   11.43341(4)\\              
Epoch (MJD)\dotfill &  49750.0\\                                      
Dispersion measure, $DM$ (cm$^{-3}$pc)\dotfill &  10.16550(3)\\
$\dot{DM}$ (cm$^{-3}$pc/yr)\dotfill &  -0.001141(7)\\
$\ddot{DM}$ (cm$^{-3}$pc/yr$^2$)\dotfill &  0.000121(3)\\
$DM^{(3)}$ (cm$^{-3}$pc/yr$^3$)\dotfill &  0.000011(1)\\
{\it DM distance}, $d$ (kpc) \dotfill & 0.6(1) \\
{\it Transverse velocity}, $V_t$ (km/s) \dotfill & 273(45)\\
{\it Kinematic correction}, $\dot{P}_k$ ($10^{-20}$) \dotfill & 8(3)\\
{\it Characteristic age}, $t_c$ (Gyr) \dotfill & 3(3)\\
{\it Surface magnetic field}, $B_s$ ($10^8$ G) \dotfill & 5(2)\\
\enddata
\tablenotetext{a}{Figures in parentheses are the formal
$1\sigma$ uncertainties in the last digits quoted.
The distance, $d$, is based on the \cite{Tay:93::} galactic electron distribution
model. 
$\dot{P}_{k}$ corrects $\dot P$ 
for accelerations due to proper motion and to vertical and
differential accelerations in the Galaxy \citep{Shk:70::,Cam:94::};
$t_c = P/2\dot{P}$,
$B_s = 3.2 \times 10^{19} (P\dot{P})^{1/2}$.} 
\end{deluxetable}

%
% TABLE 2
%

\begin{deluxetable}{lccc}
%\footnotesize
%\scriptsize
\tablewidth{390pt}
\tablecaption{Orbital and Physical Parameters of Planets \tablenotemark{a}}
\tablehead{\colhead{Parameter} & \colhead{Planet A} &  \colhead{Planet B} & 
\colhead{Planet C} }
\startdata
Projected semi-major axis, $x^0$ (ms)\dotfill  & 0.0030(1) & 1.3106(1) & 
1.4134(2) \\
Eccentricity, $e^0$\dotfill  & 0.0 & 0.0186(2) & 0.0252(2)\\
Epoch of pericenter, $T_p^0$ (MJD)\dotfill & 49765.1(2) & 49768.1(1) & 
49766.5(1) \\
Orbital period, $P_b^0$ (d)\dotfill  & 25.262(3) & 66.5419(1)  & 98.2114(2) \\
Longitude of pericenter, $\omega^0$ (deg)\dotfill & 0.0 & 250.4(6) & 108.3(5) \\
{\it Mass} ($M_{\oplus}$)\dotfill  & 0.020(2) & 4.3(2) & 3.9(2) \\
{\it Inclination, solution 1}, $i^0$ (deg) \dotfill & ... & 53(4) & 47(3) \\
{\it Inclination, solution 2}, $i^0$ (deg) \dotfill & ... & 127(4) & 133(3) \\
{\it Planet semi-major axis}, $a_{p}^0$ (AU) \dotfill & 0.19 & 0.36 & 0.46 \\
\cutinhead{Non-Keplerian Dynamical Parameters}
$\gamma_B$ (10$^{-6}$) \dotfill & & 9.2(4) & \\
$\gamma_C$ (10$^{-6}$) \dotfill & & 8.3(4) & \\
$\tau$ (deg)\dotfill & & 2.1(9) & \\
\enddata
\tablenotetext{a}{Figures in parentheses are the formal
$1\sigma$ uncertainties in the last digits quoted.}
\end{deluxetable}

\end{document}